# The Study of Media Beta Elliptical Cavities for CADS

Wen Liangjian(温良剑)[1,2]    Zhang Shenghu(张生虎)[1]    Li Yongming(李永明)[1;1)]    Wang Ruoxu(王若旭)[1]    Guo Hao(郭浩)[1]    Zhang Cong(张聪)[1]    Jia Huan(贾欢)[1]    Jiang Tiancai(蒋天才)[1]    Li Chunlong(李春龙)[1]    He Yuan(何源)[1]

[1] Institute of Modern Physics, Chinese Academy of Sciences, Lanzhou 730000, China

[2] University of Chinese Academy of Sciences, Beijing 100049, China

**Abstract:** The China Accelerator Driven Sub-critical System (CADS) is a high intensity proton facility to dispose of nuclear waste and generate electric power. CADS is based on 1.5GeV, 10mA CW superconducting (SC) linac as a driver. The high energy section of the linac is compose of two families of SC elliptical cavities which are designed for the geometrical beta 0.63 and 0.82. In this paper, the 650 MHz β=0.63 SC elliptical cavity was studied including cavity optimization, multipacting, high order modes (HOMs) and generator RF power calculation.

## 1 Introduction

China strives to develop the nuclear energy and 58 million kilowatts of the nuclear power will be reached in 2020. The nuclear waste will accumulate to 10400 tons[1]. The demand of nuclear energy will grow further with economic development. The disposal of nuclear waste and nuclear fuel shortage are increasingly serious in China. Accelerator Driven Sub-critical System (ADS) is the optimal way to dispose of nuclear waste and solve the problems of nuclear fuel shortage. CADS is promoted and constructed by Chinese Academy of Sciences (CAS), as a long-term energy strategy for China.

CADS is composed of a SC lianc, a spallation target and a nuclear reactor operating in subcritical mode. The schematic of the configuration of the SC lianc is showed in Fig.1.There are two different injectors operating in parallel, which act as a spare for each other to satisfy the strict requirement on stability in the low energy section. The injector is followed by two kinds of cavities, the spoke cavity and the elliptical cavity, which are designed to accelerate protons to 1.5GeV.

This paper is mainly concerned with the radio-frequency (RF) properties, multipacting, damping of HOMs and required generator power of the 650MHz β=0.63 elliptical cavity which accelerate protons from 180 MeV to 360MeV

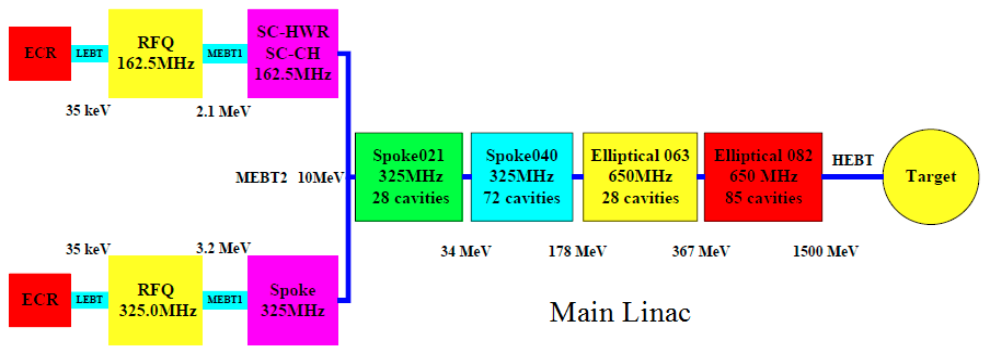

Fig.1.    The Layout of ADS linac

1) E-mail： ymli@impcas.ac.cn

## 2 Cavity RF Design

The elliptical cavity can be parameterized with the geometrical parameters shown in Fig.2. The shape of elliptical cavity is optimized for ideal electromagnetic and mechanical properties by tuning the geometrical parameters. The optimum design of an elliptical cavity is the consequence of a series of compromises among RF properties, mechanics, multipacting and HOMs damping request.

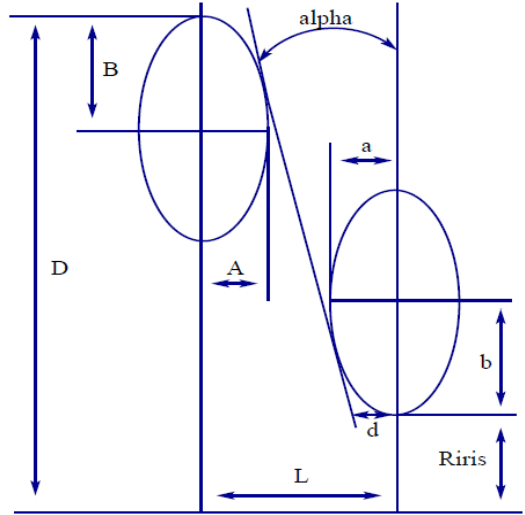

Fig.2.    the geometrical parameters of elliptical cavity

The following consideration is needed for the cavity design:

1) HOMs damping is primarily concerned in high-current SC elliptical cavity design. The larger cavity aperture reduces the interaction between the cavity and beam. And it also improves the cell-to-cell coupling to reduce potential for the trapped HOMs and beam instability. The geometry is optimized to tune frequency of the monopole HOMs away from the fundamental machine line, which could reduce power dissipation.

2) The accelerating efficiency improves by increasing the numbers of cells per cavity, which can rise the active accelerating length per meter for the whole accelerator. On the other hand, more cells per cavity not only make the transit time factor (TTF) drop faster as illustrated in Fig.3, but also make it difficult to extract and damp the HOMs. Five cells per cavity is compromise among the accelerating efficiency, the particle acceptance of TTF and damping the HOMs.

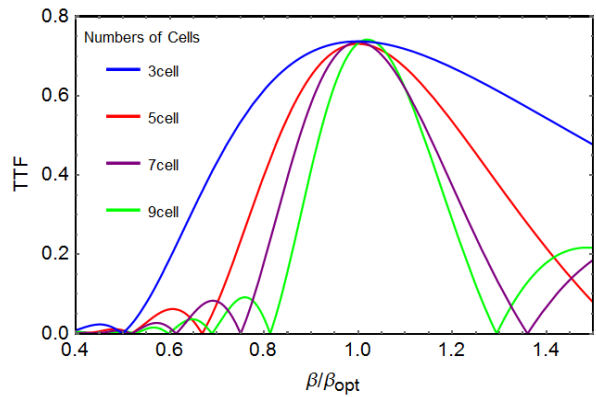

Fig.3.    The TTF versus a ratio of beam velocity β  to the optimum $\beta_{opt}$

3) The following RF properties are requested:

a) Maximize the R/Q of the accelerating mode to minimize the cavity wall dissipation

b) Minimize the $E_{pk}/E_{acc}$ and $B_{pk}/E_{acc}$ to avoid the field emission and quench at lower gradients to get a higher accelerating gradient

c) A Large cell-to-cell coupling improves field flatness and reduces potential for the trapped HOMs

A 650 MHz$\beta$=0.63 superconducting cavity was studied and designed using Superfish [2] and the 3D CST Studio Suite code[3]. The achieved cavity RF parameters are listed in Table.1. The geometry parameters are shown in Table.2. Fig.4 depicts the final $\beta$=0.63 five-cell cavity design by the 3D CST Studio Suite code.

Table.1. The RF parameters of the 650MHz $\beta$=0.63 elliptical cavity

| parameter | unit | value |
|---|---|---|
| $\beta_g$ | | 0.63 |
| frequency | MHz | 650 |
| equator diameter | mm | 394.4 |
| iris aperture | mm | 90 |
| beam pipe aperture | mm | 96 |
| cell-to-cell coupling | % | 0.9 |
| R/Q($\beta$g) | $\Omega$ | 333.5 |
| G | $\Omega$ | 192.7 |
| $E_{peak}/E_{acc}$ | | 2.34 |
| $B_{peak}/E_{acc}$ | mT/(MV/m) | 4.63 |

Table.2. The geometrical parameters of the 650MHz $\beta$=0.63 elliptical cavity

| parameters | unit | center cell | end cell |
|---|---|---|---|
| L | mm | 72.5 | 75.2 |
| iriis | mm | 90 | 96 |
| D | mm | 394.4 | 394.4 |
| A | mm | 53 | 52 |
| B | mm | 58 | 58 |
| a | mm | 16 | 15 |
| b | mm | 31 | 29 |
| $\alpha$ | ° | 3 | 4.7 |

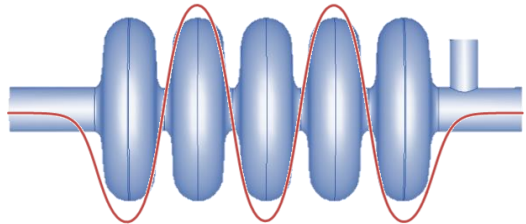

Fig.4. the final$\beta$=0.63 five-cell cavity design

# 3 Multipacting simulation

Multipacting restricts SC cavity performance and accelerating gradient enhanced since a great deal of electrons reach resonance and absorb RF power. It also leads to temperature of SC cavity rising and eventually thermal breakdown. It is crucial to optimize the shape of cavity to eliminate the unexpected multipacting barriers. Multipac 2.1[4] code was used to simulate the multipacting of the 5-cell media beta cavity. And Fig. 2 shows the enhanced counter function. The result indicates that no hard multipacting was found if the cavity can be processed by very well.

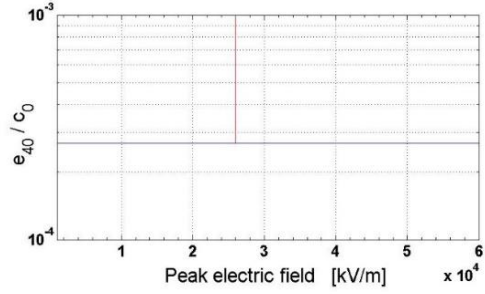

Fig.5.  Enhanced counter function- ratio of the number of particles after the 40th impact to the initial number of particles

## 4 High Order Modes (HOMs)

HOMs excited by the beam need to be analysed in detail in SC cavity. HOMs not only conversely impact on the beam and degrade the beam's quality even leading to beam loss and beam breakup. HOMs interact with cavity wall and induce the additional power dissipation and even thermal breakdown. Therefore, HOMs finding is crucial in the SC elliptical cavity design.

The effect of HOMs for beam instability in the proton linac is much lower than that in the electron linac due to the relatively high proton mass compared to that of the electron. In the case of CADS, HOMs simulation shows that the effect of longitude (mainly monopole modes) and transverse parasitic mode (mainly dipole modes) is not a big concern. And the effective transverse emittance increase induced by HOMs is lower than the tolerable limit even for the external $Q=10^8$ and the beam current =100mA[5]. If the frequency of HOMs are far away from the machine line and R/Q are smaller, the longitudinal beam instability induced by HOMs has little effect [6]. Therefore, the frequency of HOMs is enough far away from the machine in consideration of the variation of the HOMs' frequency due to imperfect manufacture.

The power dissipation induced by TM-monopoles in cavity wall is much larger than that induced by multipoles such as dipoles and quadrupole. Hence the calculation of the power dissipation by HOMs mainly concerns TM-monopoles [7]. If the voltage of HOMs excited by beam in SC cavity reached an equilibrium state in CW operation, the actual power dissipated by mode n in cavity wall can be directly calculated by the following formula which does not take consideration of the beam noise.

$$P_{c,n} = \frac{|V_n|^2}{(R/Q)_n(\beta)Q_{0,n}} \qquad (1)$$

Here, $V_n$ is longitudinal cavity voltage of the mode n of HOMs excited by beam.

The Fig.6 shows the actual power dissipated induced by HOMs in the media beta cavity walls in the parameter space of $Q_{ex}$ and frequency. We can conclude that at resonance the dissipated power exponentially increases with $Q_{ex}$ and reaches to 100 W/Ω at $Q_{ex} = 10^8$ and drops faster with $Q_{ex}$ when the frequency is far away from the machine line from Fig.6 (a). Fig.6 (b) illustrates the dissipated power in parameter of frequency among the machine line with the different $Q_{ex}$. The bandwidth of the dissipated power at resonance with $Q_{ex} = 10^8$ is much smaller than that with $Q_{ex} = 10^5$. The probability of HOMs at resonance with $Q_{ex}=10^8$ are much smaller than that with $10^5$.

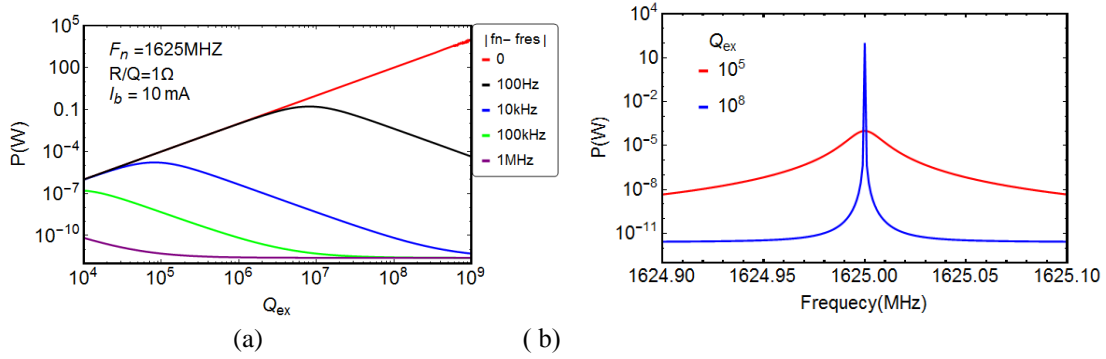

Fig.6.    HOM power dissipation in cavity wall versus Qex (a) and frequency (b)

Frequency, $Q_{ex}$ and R/Q of each HOM can be simulated with the 3D CST Studio Suite code.Table.3 depicts the frequency and $Q_{ex}$ of TM-monopoles. And Fig.7 shows the maximize R/Q from 180 MeV to 360 MeV of TM-monopoles. R/Q of monopoles except for the fundamental modes is much small compared to the accelerating mode and the $Q_{ex}$ of most monopole modes is about $\sim 10^8$. The theoretical dissipated power of Monopole is about $\sim 10^{-10}$ without considering the beam noise and variation of HOMs' frequency in actual operation.

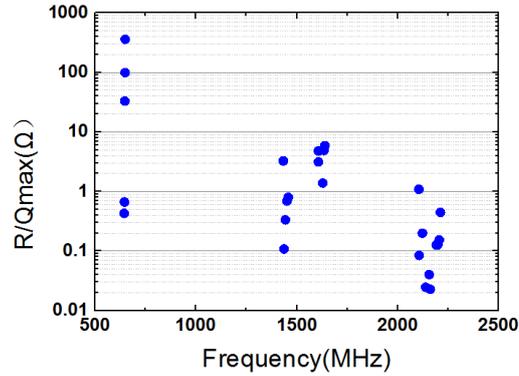

Fig.7.    The maximize R/Q from 180 MeV to 360 MeV of TM-monopoles. R/Q of monopoles

Table.3. The frequency and $Q_{ex}$ of TM-monopoles of the elliptical cavity

| Mode | Frequency/ MHz | off resonance/MHz | Qex | Mode | Frequency/ MHz | off resonance/MHz | Qex |
|---|---|---|---|---|---|---|---|
| TM020,1/5π | 1432.95 | 29.55 | 7.30E+08 | TM011,1/5π | 1606.90 | 18.10 | 1.65E+07 |
| TM020,2/5π | 1436.52 | 25.98 | 2.86E+08 | TM011,2/5π | 1606.97 | 18.03 | 4.92E+07 |
| TM020,3/5π | 1442.81 | 19.69 | 2.53E+08 | TM011,3/5π | 1628.56 | 3.56 | 1.28E+09 |
| TM020,4/5π | 1450.60 | 11.90 | 3.69E+08 | TM011,4/5π | 1634.21 | 9.21 | 1.10E+09 |
| TM020,5/5π | 1457.31 | 5.19 | 1.13E+09 | TM011,5/5π | 1638.85 | 13.85 | 3.20E+09 |
| TM021,1/5π | 2104.32 | 8.18 | 3.56E+05 | TM012,1/5π | 2160.58 | 48.08 | 1.84E+09 |
| TM021,2/5π | 2105.81 | 6.69 | 4.21E+05 | TM012,2/5π | 2190.58 | 78.08 | 1.78E+02 |
| TM021,3/5π | 2121.64 | 9.14 | 3.07E+06 | TM012,3/5π | 2197.24 | 77.76 | 6.16E+04 |
| TM021,4/5π | 2138.55 | 26.05 | 1.60E+02 | TM012,4/5π | 2205.00 | 70.00 | 6.11E+12 |
| TM021,5/5π | 2155.22 | 42.72 | 6.32E+07 | TM012,5/5π | 2211.01 | 63.99 | 4.33E+02 |

## 5 Required Generator Power

The 650 MHz β=0.63 SC cavity is design to operate in the accelerating gradient ($E_{acc}$) of 15 MV/m. The optimum eternal Q need to be considered to minimize the generator power of the input power coupler. In fact, the practical requested power need to consider the effect of frequency detuning and the accelerating gradient upgrading in the cavity design. The generator power is determined by the following equation [8] at a given accelerator field.

$$P_g = \frac{V_c^2}{R_{sh}} \frac{1}{4\beta} \{(1+\beta+b)^2 + [(1+\beta)\tan\psi - b\tan\phi]^2\} \quad (2)$$

And

$$b = \frac{R_{sh} i_0 \cos\phi}{V_c}$$

$$\tan\psi = -2\frac{Q_0}{1+\beta}\frac{\Delta\omega}{\omega}$$

Here, $R_{sh}$ is the shunt impedance, $V_c$ is the voltage in the cavity, $i_0$ is the beam dc current, $\phi$ is the detuning angle and $\psi$ is the beam phase.

The RF power needed for the 650 MHz β=0.63 SC cavity versus the external Q at $E_{acc}$ = 15MV/m is illustrated in Fig.8. If a conservative detuning of 25Hz is assumed, we could get the optimum external Q ranging from 2.9×10⁶ to 3.3×10⁶ and the moderate generator power is 120 KW. Fig.9 and Fig.10 depict the minimize Generator Power and the optimum eternal Q versus the Eacc. The optimum generator power would be 150 KW which could make the accelerating gradient reach to 20 MV/m.

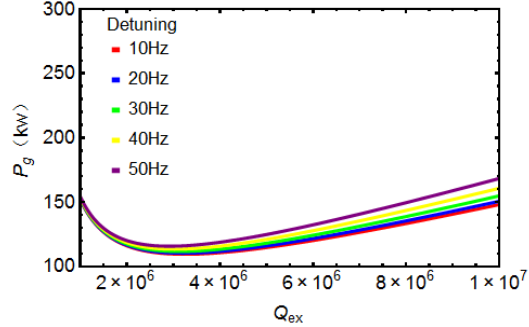

Fig.8.    The RF power versus the external Q
at $E_{acc}$ = 15MV/m

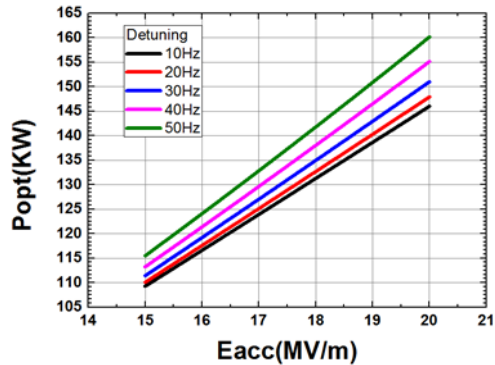

Fig.9.    the minimize Generator Power
versus the Eacc

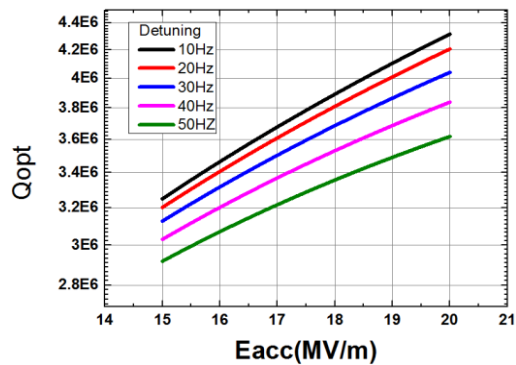

Fig.10.   the optimum eternal Q
versus the Eacc

## 6 Summary

The 650 MHz β=0.63 SC elliptical cavity is designed for the CADS linac, accelerating proton beams from 180 MeV to 360 MeV. The RF and geometry parameters meets the request for CADS linac. The multipacting was checked by the Multipac 2.1 and no hard multipacting barriers are found in the cavity. The problems induced by HOMs are theoretically not a main concern in the β=0.63 SC elliptical cavity. The optimum external Q ranges from $2.9 \times 10^6$ to $3.3 \times 10^6$ and the moderate generator power is 120 KW at the designed gradient $E_{acc}$=15MV/m.